\documentclass[12pt]{article}

\usepackage{amsmath,amssymb,amsfonts}

\newcommand{\RR}{{\mathbb R}}

\newcommand{\ZZ}{{\mathbb Z}}

\newcommand{\ra}{\rightarrow}
\renewcommand{\k}{{\mathsf k}}
\renewcommand{\v}{{\mathsf v}}
\newcommand{\pM}{{\partial M}}

\newcommand{\cO}{{\mathcal O}}
\newcommand{\hX}{{\hat X}}
\newcommand{\hg}{{\hat g}}
\newcommand{\hB}{{\hat B}}

\newcommand{\cL}{{\mathcal L}}
\newcommand{\eps}{{\epsilon}}
\newcommand{\Z}{{\mathcal Z}}
\newcommand{\bs}{{\backslash}}

\newcommand{\ch}{{\rm ch}}
\newcommand{\hPhi}{{\hat\Phi}}
\newcommand{\p}{{\mathsf p}}

\newcommand{\cP}{{\mathcal P}}
\newcommand{\hL}{{\hat L}}

\newcommand{\hA}{{\hat A}}
\newcommand{\he}{{\hat e}}
\newcommand{\htheta}{{\hat \theta}}
\newcommand{\hF}{{\hat F}}
\newcommand{\hT}{{\hat T}}
\newcommand{\htau}{{\hat\tau}}

\DeclareMathOperator{\Harm}{Harm}

\begin{document}

\begin{titlepage}

\title{Abelian duality, walls and boundary conditions in diverse dimensions}
\author{Anton Kapustin \\
{\it\small California Institute of Technology} \\{\it\small Pasadena, CA 91125, United
States}
\and Mikhail Tikhonov \\ {\it\small D\'epartement de physique,
\'{E}cole Normale Sup\'{e}rieure,}\\{\it\small 75231 Paris Cedex 05,
France}}

\maketitle

\begin{abstract}
We systematically apply the formalism of duality walls to study the
action of duality transformations on boundary conditions and local
and nonlocal operators in two, three, and four-dimensional free
field theories. In particular, we construct a large class of
D-branes for two-dimensional sigma-models with toroidal targets and
determine the action of the T-duality group on it. It is manifest in
this formalism that T-duality transformations on D-branes are given
by a differential-geometric version of the Fourier-Mukai transform.

\end{abstract}

\end{titlepage}

\section{Introduction}

It has been realized recently (see e.~g. \cite{KW,GW,Bi1,Bi2,Bi3})
that the most economic description of duality transformations in
field theory is by means of codimension-1 defects (walls). Any
equivalence between quantum field theories A and B gives rise to a
domain wall which separates a domain described by the field theory A
and a domain described by the field theory B. We will refer to these
domains as phase A and phase B. While there may be many different
domain walls between these phases, the wall corresponding to a
duality transformation has a very special property: it is
topological, in the sense that correlation functions are unchanged
if one deforms the location of the wall without crossing the
insertion points of any operators. Given such a wall, one can
determine how the duality acts on any local operator $\cO$ in the
theory A: one simply considers a ``composite'' made of $\cO$ and the
wall wrapping a small sphere centered at the insertion point of
$\cO$, so that the interior of the sphere is in phase A, while the
exterior is in phase B. In the limit when the radius of the sphere
goes to zero, this composite defines a local operator in the theory
B. Similarly, one can determine how nonlocal operators and boundary
conditions transform under duality transformations.

In this paper we systematically apply these ideas to abelian
dualities in two, three, and four dimensions. In 4d we study duality
transformations for the free $U(1)$ gauge theory. There is a
substantial overlap here with a recent work of Gaiotto and
Witten~\cite{GW}. For example, Gaiotto and Witten explain how to
find the electric-magnetic dual of an arbitrary boundary condition
in 4d $U(1)$ gauge theory. Our contribution is to provide some
examples of using the wall formalism to determine the mapping of
nonlocal operators under the $SL(2,\ZZ)$ duality group. For example,
we explain how to find the dual of the Chern-Simons operator in the
4d $U(1)$ gauge theory, something which is difficult to do using the
traditional methods.

In the two-dimensional case, we show how the formalism of duality
walls enables us to define the action of T-duality on a large class
of branes on tori in a unform way. For example, we show by a simple
manipulation of the path-integral that the T-dual of a vector bundle
on $T^2$ of rank $r$ and first Chern class $k$ is a vector bundle on
the dual torus of rank $k$ and first Chern class $-r$. Boundary
degrees of freedom describing the Chan-Paton labels appear naturally
in this derivation. This is an improvement over the traditional
argument which uses step-wise T-duality (dualizing circles one by
one). More generally, we show that for our class of D-branes
T-duality acts as a differential-geometric version of the
Fourier-Mukai transform. Domain walls corresponding to T-dualities
have been recently discussed in \cite{Bi3}. The connection between
T-duality and Fourier-Mukai transform has also been noted in the
literature, but usually in the context of supersymmetric
sigma-models and supersymmetric branes. We do not make use of
supersymmetry in this paper.

In the three-dimensional case the abelian duality relates a $U(1)$
gauge theory with a Maxwell action and a massless scalar field. We
describe the wall corresponding to this duality and use it to deduce
how the duality acts on some boundary conditions and operators.
Somewhat unexpectedly, to define the duality wall precisely one needs
to make use of the notion of a gerbe connection (also known as a
B-field). We also discuss how the introduction of the Chern-Simons
term in the gauge theory destroys the duality.

M.~T. would like to thank California Institute of Technology for
hospitality during the initial stages of this work. A.~K. was
supported in part by the DOE grant DE-FG03-92-ER40701.

\section{Abelian duality in 4d}

\subsection{Boundary conditions for a $U(1)$ gauge theory}

We will study a $U(1)$ gauge theory on a Riemannian 4-manifold $M$
with an action
$$
S=\frac{1}{2e^2}\int_M F\wedge *F+\frac{i\theta}{8\pi^2}\int_M
F\wedge F.
$$
Here $F=dA$ is the curvature of the $U(1)$ gauge field $A$. The
simplest boundary condition for the gauge field is to require the
restriction of $A$ to $\pM$ to be trivial (i.~e. gauge-equivalent to
zero). More generally, we can require $A\vert_\pM$ to be a flat, but
not necessarily trivial (fixed) connection. The curvature then
satisfies
\begin{equation}\label{PEC}
F\vert_\pM=0.
\end{equation}
If the boundary contains a time-like direction, then this condition
means, in three-dimensional terms, that the electric field is normal
to the boundary, while the magnetic field is tangent to it. Thus it
corresponds to a perfect electric conductor (PEC).

Another simple condition is the free boundary condition which allows
$A\vert_\pM$ to be unconstrained. The vanishing of the boundary part
of the variation of the action then gives the following boundary
condition on the curvature:
\begin{equation}\label{PMC}
\frac{1}{e^2} *F\vert_\pM+\frac{i\theta}{4\pi^2} F\vert_\pM=0.
\end{equation}
For $\theta=0$ it means that the electric field is tangent to the
(space-like) boundary, while the magnetic field is normal to it.
This corresponds to a magnetic with an infinitely large magnetic
permeability. Since the perfect electric conductor is formally
equivalent to a dielectric with an infinite electric permeability,
it is natural to regard the boundary condition (\ref{PMC}) as
corresponding to a ``perfect magnetic conductor'' (PMC).

This terminology is also natural because the PEC and PMC boundary
conditions are related by electric-magnetic duality. This is rather
obvious for $\theta=0$ since electric-magnetic duality for
$\theta=0$ exchanges electric and magnetic fields. It is also true
for arbitrary $\theta$ as we will see below.

As explained above, on the quantum level the PEC condition has
moduli which are described by a flat $U(1)$ connection on $\pM$.
Neglecting the torsion phenomena, the moduli space of such
connections is a torus
$$
H^1(\pM,\RR)/2\pi H^1(\pM,\ZZ).
$$
There are also moduli in the PMC case. We can add to the action a
boundary term
$$
\frac{i}{2\pi}\int_\pM F\wedge \lambda,
$$
where $\lambda$ is a closed 1-form on $\pM$.\footnote{The form
$\lambda$ is taken to be closed so that the extra term does not
contribute to the boundary variation of the action and therefore
does not affect the boundary condition (\ref{PMC}).} Since the
periods of $F$ are integers modulo $2\pi$, the action is invariant
under
$$
\lambda\mapsto\lambda+2\pi\alpha,
$$
where $\alpha$ is a closed 1-form on $\pM$ with integral periods.
The set of closed 1-forms modulo such transformations is again the
torus
$$
H^1(\pM,\RR)/2\pi H^1(\pM,\ZZ).
$$
We will see below that PEC and PMC moduli are exchanged by
electric-magnetic duality.

To realize the full duality group $SL(2,\ZZ)$ one has to consider
more general boundary conditions. In fact, to realize $SL(2,\ZZ)$
quantum-mechanically one has to allow boundary degrees of freedom.
These degrees of freedom are topological, in the sense that in the
limit $e\ra 0$ they are described by a 3d topological gauge theory
on $\pM$.

A natural 3d topological field theory which can be coupled to a
$U(1)$ gauge theory is abelian Chern-Simons theory \cite{Witten:ST}.
Its action is
$$
\sum_{jl}\frac{i\k_{jl}}{4\pi} \int_\pM a^j da^l.
$$
where the indices $j,l$ run from $1$ to $n$ and $a^j$ is a
connection 1-form on a $U(1)$ bundle over $\partial M$. The matrix
$\k_{jl}$ is symmetric and integral; this ensures the invariance of
the path-integral under large gauge transformations. This 3d theory
has conserved currents
$$
J^l=* \frac{1}{2\pi} d a^l,\quad l=1,\ldots,n.
$$
The corresponding charges are quantized:
$$
\int_\Sigma * J^l \in \ZZ,
$$
where $\Sigma$ is a 2-cycle in $\partial M$. These charges are
simply fluxes of the gauge fields $a^j$.

To couple such a topological gauge theory to a 4d $U(1)$ gauge
theory one needs to choose a $U(1)$ current (which is a linear
combination of $J_l$ with integer coefficients) and add to the
action a boundary term
\begin{equation}\label{bbcoupling4d}
\frac{i}{2\pi}\int_\pM A \sum_l \v_l da^l
\end{equation}
The row-vector $\v_l$ is integral.

Ordinarily, one requires the matrix $\k_{jl}$ to be nondegenerate,
so that after gauge-fixing there is a well-defined propagator.
However, in the presence of a coupling to bulk fields one can relax
this assumption. Indeed, if $\k$ is degenerate, this means that
certain fields simply do not appear in the Chern-Simons part of the
Lagrangian. However, the linear combination $\v_l a^l$ appears in
the term (\ref{bbcoupling4d}). If the kernel of $\k$ is
one-dimensional, and the row-vector $\v$ does not annihilate it,
then the field which does not appear in the Chern-Simons part of the
action is simply a Lagrange multiplier integration over which forces
an integer multiple of $A$ to be trivial on the boundary. For
example, the PEC condition corresponds to $n=1$, $\k=0$, and $\v=1$.
Thus we will assume that $\k$ is either nondegenerate, or has a
one-dimensional kernel which is not annihilated by $\v$. We will
call such a boundary condition nondegenerate. If we regard $\v$ as a linear operator
$\v\colon\RR^n\mapsto \RR$, then the non-degeneracy condition can be compactly written as
$\ker \k\cap\ker \v=0$.

Finally, one may also add to the boundary action a Chern-Simons term
for the restriction of the bulk gauge field:
$$
\frac{i \p}{4\pi}\int_\pM A\,dA .
$$
Gauge-invariance requires the coefficient $\p$ to be integral. The
most general boundary condition we are going to consider is
described by an integral square matrix $\k$, an integral vector $\v$
and an integer $\p$. The $SL(2,\ZZ)$ duality group acts on this set
of boundary conditions, as first explained in \cite{Witten:ST}.

On the classical level, we can integrate out the fields $a_l$ and
get an ordinary boundary condition for $A$. However, in general this
leads to a boundary effective action for $A$ which looks like
Chern-Simons action with a fractional coefficient. This is
inconsistent on the quantum level.

In particular, this means that a consistent treatment of a ``perfect
dyonic conductor'' boundary condition requires an introduction of
topological degrees of freedom living on the boundary.

\subsection{$SL(2,\ZZ)$ action on boundary conditions}

It follows from the results of \cite{Witten:ST} that the set of
boundary conditions introduced above is closed with respect to the
action of the group of duality transformations. In this section we
review how this comes about following the approach of \cite{GW}; in
particular, we show that the moduli of PEC and PMC boundary
conditions are mapped into each other by electric-magnetic duality.

The duality group of the $U(1)$ gauge theory is $SL(2,\ZZ)$. Its
action on the parameters of the theory is most easily described by
introducing
$$
\tau=\frac{\theta}{2\pi}+\frac{2\pi i}{e^2}.
$$
Then an element of $SL(2,\ZZ)$
$$
g=\begin{pmatrix} a & b\\ c & d\end{pmatrix},\quad
a,b,c,d\in\ZZ,\quad ad-bc=1
$$
acts on $\tau$ by
$$
\tau\mapsto \hat\tau=\frac{a\tau+b}{c\tau+d}.
$$
The same transformation maps a dyon of electric charge $n$ and
magnetic charge $m$ to a dyon whose charges are
$$
\begin{pmatrix} {\hat n} \\ {\hat m}\end{pmatrix}=g^{-1}\begin{pmatrix} n \\ m
\end{pmatrix}.
$$
As explained in \cite{GW}, any element $g\in SL(2,\ZZ)$ gives rise
to a defect of codimension $1$ in the 4d theory which separates two
$U(1)$ gauge theories related by a duality transformation. The
action of $g$ on a boundary condition is obtained by fusing the
defect with the boundary. For brevity we will call the nonlocal
operator corresponding to a defect of codimension $1$ the wall
operator. Operators corresponding to defects of codimension $2$ and
$3$ will be called surface operators and line operators,
respectively. For example, the Wilson loop is a line operator.

The group $SL(2,\ZZ)$ is generated by elements
$$
T=\begin{pmatrix} 1 & 1 \\ 0 & 1\end{pmatrix},\quad
S=\begin{pmatrix} 0 & 1 \\ -1 & 0\end{pmatrix},
$$
so it is sufficient to determine how these elements act on boundary
degrees of freedom.

The $T$-transformation (i.~e. the shift $\theta\ra\theta+2\pi$) is
realized by the wall operator
\begin{equation}\label{Twall}
\exp\left(-\frac{i}{4\pi}\int_W A\,dA\right)
\end{equation}
Here $W$ is a 3-dimensional submanifold in $M$. This wall is
topological because the expression (\ref{Twall}) is independent of
the metric, and therefore does not affect the stress-energy tensor.
To illustrate that this is indeed the wall which corresponds to the
$T$-transformation, consider its action on a 't~Hooft loop operator
localized on a circle $\gamma$ embedded into the interior of $M$.
Let $W$ be a wall which is the boundary of a tubular neighborhood of
$\gamma$. The 't~Hooft operator is defined by the condition that the
curvature $F$ is singular at $\gamma$ so that the integral of $F$
over a small 2-sphere linking $\gamma$ is $2\pi m$. Here $m$ is the
magnetic charge of the 't~Hooft loop operator. Then the wall
operator becomes
$$
\exp\left(-i m \int_\gamma A\right).
$$
Thus the 't~Hooft loop operator with magnetic charge $m$ is
transformed into a Wilson-'t~Hooft loop operator with electric
charge $-m$ and magnetic charge $m$. This is the expected result.

Next we consider the wall operator implementing the
$S$-transformation, i.~e. the electric-magnetic duality. As explained
in \cite{GW}, it is defined as follows. Suppose $W$ splits $M$ into
two disjoint pieces which we call $M_-$ and $M_+$. We will choose
the orientation of $W$ so that it agrees with the one induced from
$M_-$ and disagrees with the one induced from $M_+$. Let $A$ and
$\hA$ be the gauge fields living on $M_-$ and $M_+$. We do not put
any constraints on their restrictions to $W$ and add the following
term to the action:
$$
\frac{i}{2\pi}\int_W A\,d\hA.
$$
We denote the gauge couplings and theta-angles of the gauge theories
on $M_-$ and $M_+$ by $e,\theta$ and $\he,\htheta$, respectively.

To see that this wall operator implements the $S$ transformation, we
may do the following. Varying the action and requiring the part of the
variation localized on $W$ to vanish, we find the matching
of the fields across the boundary:
\begin{equation}\label{eq:matchingCond}
\begin{aligned}
\hF\vert_W &=\frac{2\pi
i}{e^2}*F\vert_W - \frac{\theta}{2\pi} F\vert_W,\\
F\vert_W & =-\frac{2\pi i}{\he^2}*\hF\vert_W + \frac{\htheta}{2\pi}
\hF\vert_W .
\end{aligned}
\end{equation}
Even if we do not impose any relation between $e,\theta$ and
$\he,\htheta$, these formulas define a wall between the two gauge
theories. However, this wall will not be a topological wall, in
general. For a topological wall, all components of the stress-energy
tensors $T$ and $\hT$ must match on $W$. The stress-energy tensor is
given by
$$T_{\mu\nu}=\frac1{e^2}\left(F_{\mu\alpha}F^\alpha_{\,\,\nu}+
\frac14g_{\mu\nu}F_{\alpha\beta}F^{\alpha\beta}\right).$$ Using the
matching conditions~\eqref{eq:matchingCond} one can easily check
that $T=\hT$ on the wall if and only if one has the relationship
\begin{equation}\label{eq:matchingCouplings}
\htau=-1/\tau,
\end{equation}
as expected.

Now let us describe the action of $S$ and $T$ wall operators on
boundary conditions. When the $T$-wall merges with a boundary, the
boundary action is modified by a term
$$
\frac{i}{4\pi}\int_\pM A\,dA
$$
This is the $T$ transformation as defined in \cite{Witten:ST}. Thus
the matrix $\k$ and the row-vector $\v$ are unchanged, while the
integer $\p$ is shifted:
$$
\k\mapsto \k,\quad \v\mapsto \v,\quad \p\mapsto \p+1.
$$

The effect of the $S$-transformation is more interesting. We assume
that the wall is oriented so that $M_-$ is the gap between the
boundary and the wall. Then in the limit when the gap disappears $A$
becomes a new boundary gauge field. The bulk action for $A$ can be
dropped in this limit, and the new boundary action is
$$
\sum_{jl}\frac{i\k_{jl}}{4\pi} \int_\pM a^j
da^l+\frac{i}{2\pi}\int_\pM a^{n+1} \sum_{l=1}^n \v_l da^l +
\frac{i}{2\pi}\int_\pM a^{n+1} d\hA,
$$
where we renamed $A=a^{n+1}$. One can describe the effect of the $S$
transformation by saying that the external gauge field $A$ has been
replaced by a boundary gauge field $a^{n+1}$, and the new external
gauge field $A=A_+$ is coupled to the boundary current
$$
J^{n+1}=\frac{1}{2\pi} da^{n+1}.
$$
This is precisely the $S$ transformation as defined in
\cite{Witten:ST}. It increases the number of the boundary fields by
one and transforms the data $(\k,\v,\p)$ as follows:
\begin{equation}\label{eq:dataTransform4d}
\left\{
\begin{aligned}
\k &\mapsto \tilde \k =\begin{pmatrix} \k & \v^t \\ \v & \p \end{pmatrix},\\
\v &\mapsto \tilde \v =(0,\ldots,0,1),\\
\p &\mapsto \tilde \p = 0.
\end{aligned}\right.
\end{equation}

It is easy to see that the new boundary condition is non-degenerate
in the sense defined above, i.~e.~that if $\ker \k\cap\ker\v=0$,
then $\ker \tilde\k\cap\ker\tilde\v=0$. Indeed, any $\tilde u$ that
lies in $\ker\tilde\v$ is of the form
$\left(\begin{smallmatrix}{\vec u}\\0\end{smallmatrix}\right)$ for
an arbitrary $n$-vector $\vec u$. But then
$$
\tilde\k\cdot\tilde u=\begin{pmatrix}\k \vec u\\\vec \v\cdot \vec u\end{pmatrix}=0\quad\Leftrightarrow\quad
\vec u=0,
$$
which is what we had to show.

To illustrate this general procedure of dualizing boundary conditions, let us show that PEC and PMC
boundary conditions are indeed dual to each other. The PEC boundary
condition can be viewed as a special case where $n=1$, $\k=0$, and
the boundary action is
$$
\frac{i}{2\pi}\int_\pM A\,da^1.
$$
Integration over $a^1$ yields a gauge-invariant delta-functional
which sets $A$ to the trivial connection. $S$ transformation maps it
to
$$
\frac{i}{2\pi}\int_\pM a^2 da^1 + \frac{i}{2\pi} \int_\pM a^2 dA.
$$
Integration over $a^2$ gives a delta-functional which sets $a^1=-A$.
Integration over $a^1$ then gives a theory with zero boundary action
and no boundary fields. This is the PMC boundary condition.

The PMC boundary condition is a special case with $n=0$ and trivial
boundary action. The $S$ transformation introduces a single boundary
gauge field $a^1$ which couples to $A$ via
$$
\frac{i}{2\pi}\int_\pM a^1 dA.
$$
This is precisely the PEC boundary condition.

In the same way we can show that $S$ duality maps the moduli of the
PMC boundary condition to the moduli of the PEC boundary condition.
For example, the $S$ transformation maps the generalized PMC
boundary condition to
$$
\frac{i}{2\pi}\int_\pM (a^1 dA + a^1 d\lambda)
$$
Integration over $a^1$ then gives a delta-functional which sets
$A=-\lambda$ up to a gauge transformation. This is the generalized
PEC boundary condition.

Applying more general elements of $SL(2,\ZZ)$ to the PEC and PMC
boundary conditions we get boundary conditions corresponding to
``perfect dyonic conductors'' with dyons having relatively-prime
electric and magnetic charges $(n,m)$. For example, applying the
transformation $T^k$ to the PMC boundary condition, we find that a
dyonic superconductor with dyons of charge $(-k,1)$ can be modeled
by a boundary Chern-Simons term at level $k$. Applying to this
boundary condition the transformation $S^{-1}$, we find the boundary
condition for the dyonic superconductor with dyons of charge
$(1,k)$. It involves a boundary gauge field $a$ with an action
$$
\frac{ik}{4\pi}\int_\pM a\,da - \frac{i}{2\pi} \int_\pM a\,dA .
$$

\subsection{The action of duality on operators}

We can use the $S$-wall to determine the action of electric-magnetic
duality on various operators, both local and nonlocal. Let us give
some examples.
\subsubsection{Wilson and 't~Hooft loops}
\label{sec:dualizingWilson4d}

We have already shown that the $S$-transformation maps the 't~Hooft loop
operator to a Wilson loop operator. Now let us show that the
$S$-transformation maps the Wilson loop operator to a 't~Hooft loop
operator.
Consider the Wilson operator
$$
W_n(\gamma)=\exp\left(in\int_\gamma A\right),
$$
where $\gamma$ is a circle embedded into $M$, and $n\in\ZZ$ is the
electric charge. Here is an outline of the argument that will allow us to calculate its dual: we
begin by regularizing this operator and introducing a parameter $\epsilon$. We will then show that
in the dual theory, only specific topological sectors contribute to the path integral, separate the
contribution of the topological representative and of the topologically trivial part of the dual field,
and find that this will allow us to calculate the result of sending $\epsilon$ to zero.

Let $\Z_\gamma^\eps$ be a tubular neighborhood of $\gamma$, with
$\eps$ being its ``thickness''. To regularize the Wilson operator,
we will need a closed and coclosed two-form $\Omega_n$ on
$M\backslash \Z_\gamma^\eps$ whose periods are integral multiples of
$2\pi$ and which satisfies the following two conditions. First, it
must satisfy
$$
* \Omega_n\vert_{\partial \Z_\gamma^\eps}=0.
$$
To explain the second condition, let us identify $\Z_\gamma^\eps$
with $S^1\times B^3_\eps$, where $B^3_\eps\subset\RR^3$ is a ball of
radius $\eps$. It is assumed that $\gamma$ has the form $S^1\times
\{0\}$, where $\{0\}$ is the center of $B^3_\eps$. For any $p\in
S^1$ we may consider the 2-sphere $S^2_\gamma=\{p\}\times \partial
B^3$ as a submanifold of $M$. Clearly, all such 2-spheres are
homologous, so the integral of $\Omega_n$ over $S^2_\gamma$ does not
depend on the choice of $p\in S^1$. We require the integral to be
equal to $2\pi n$.

If $\gamma$ is homologically trivial, then one can define an
integer-valued linking number between $\gamma$ and any 2-cycle $c$
in the complement of $\gamma$. This is done as follows: if $D$ is a
2-chain whose boundary is $\gamma$, then the linking number is the
algebraic number of intersection points of $D$ and $c$. One can show
that this number does not depend on the choice of $D$ and does
change if one replaces $c$ by a homologous 2-cycle. It is easy to
see that $\gamma$ and $S^2_\gamma$ have linking number one, which
means that $S^2_\gamma$ is homologically nontrivial when regarded as
a cycle in $M\backslash \gamma .$

Let us show that the form $\Omega_n$ indeed exists, beginning with
the case when the loop $\gamma$ is homologically trivial. The
question of existence of a harmonic representation of a de Rham
cohomology class on a manifold with boundary is classical and is
discussed for example in~\cite{Gilkey, GLP} (see also
\cite{Zucchini}); in particular, one result is that if we define
``absolute'' boundary conditions for a form $\alpha$ as
$*\alpha|_{\partial M}=0,
*d\alpha|_{\partial M}=0$, one has the isomorphism $H^p(M,\RR)\simeq
\Harm^p_{\mathrm{abs}}(M)$. This follows from the following analog
of the Hodge decomposition theorem on a manifold with boundary:
$$
\Omega^{p}(M)=d\Omega^{p-1}(M)\oplus d^*\Omega^{p+1}_{\mathrm{tan}}(M)\oplus \Harm^p_{\mathrm{abs}}(M),
$$
where $\Omega_{\mathrm{tan}}$ denotes forms satisfying the ``tangential'' boundary condition
$*\alpha|_{\partial M}=0$.

This result guarantees the existence of a harmonic representative in
each absolute cohomology class. A suitable cohomology class,
however, exists only under a certain condition. In the case of a
single Wilson operator, this condition is the requirement that
$S^2_\gamma$ should not be contractible in $M\setminus \Z_\epsilon$,
otherwise the constraint $\int_{S^2_\gamma}\Omega_n=2\pi n$ would
contradict the fact that $\Omega_n$ is closed. This condition is
satisfied if the linking number of $S^2_\gamma$ and $\gamma$ is one.

We may consider a more general case when several Wilson loops
$\gamma_1,\ldots,\gamma_K$ with charges $n_1,\ldots,n_K$ are present
simultaneously. We claim that a form $\Omega_{\{n_i\}}$ that
integrates to  $n_i$ on the 2-spheres $S^2_{\gamma_i}$ exists if and
only if
\begin{equation}\label{topcond}
\sum_i n_i[\gamma_i]=0,
\end{equation}
where $[\gamma_i]$ denotes the homology class of $\gamma_i$. Only in
this case will we be able to find the $S$-dual operator
corresponding to such a combination of Wilson loops. This, however,
is not a problem since if this homological condition is not
satisfied, the path integral calculating the corresponding Wilson
loop correlator vanishes. (The action and the integration measure
are invariant under shifts of the gauge field by a flat connection,
so any operator insertion must either be invariant as well, or yield
a zero correlator. The invariance condition is precisely
\eqref{topcond}).

Let us now prove the statement we made about the existence of
$\Omega_{\{n_i\}}$. If the topological condition on $\{\gamma_i\}$
is satisfied, there exists a 2-chain $D$ whose boundary is $n_1$
copies of $\gamma_1$, $n_2$ copies of $\gamma_2$, etc. It is
therefore a relative 2-cycle in what we will call the ``bulk'':
$M\setminus\left(\bigcup_i {\mathcal Z}^\epsilon_{\gamma_i}\right)$.
Consider the Poincare-dual class in the absolute cohomology of the
bulk. This Poincare-dual class has a harmonic de Rham representative
with the desired boundary conditions (by the above-mentioned Hodge
decomposition theorem on a manifold with boundary). This is the form
$\Omega_{\{n_i\}}$ we need. Although we do not really need the
converse statement, let us sketch its proof. If $\Omega_{\{n_i\}}$
exists, then it defines an absolute de Rham cohomology class whose
Poincare-dual relative homology class can be realized by a 2-chain
whose boundary lies on $\partial \bigcup_i {\mathcal
Z}^\epsilon_{\gamma_i}$. It is easy to see that in fact this
boundary is homologous to the sum
$$
\sum_i n_i[\gamma_i].
$$
Applying the long exact sequence of a pair, we conclude that this
sum must be homologically trivial when regarded as a class in the
bulk.

{}From now on, to simplify the notation, we will return to the case
of a single Wilson loop $\gamma$; the generalization of the argument
to multiple loops is straightforward. Having found $\Omega_n$, we
define the regularized Wilson loop by
$$
W^\epsilon_n(\gamma)=\exp\left(\frac{i}{2\pi}\int_{\partial\Z_\eps} A\wedge
\Omega_n\right).
$$
To see that this is indeed a regularization of the Wilson loop, note
that in the limit $\eps\ra 0$ we may approximate $\gamma$ by a
straight line in $\RR^4$, and then in spherical coordinates
$(t,r,\theta,\phi)$ with $t$ running along $\gamma$ the form
$\Omega_n$ is well approximated by $2\pi n$ times the unit volume
form $\Omega$ on the two-sphere $(t=t_0,\; r=\epsilon)$. Then the
integral in the exponential becomes
$$
in \int_{r=\eps} A_t\, d\Omega\, dt
$$
which is obviously a regularization of $(in\int_\gamma A)$.

Having regularized the Wilson loop, we place the duality wall on the
boundary of $\Z_\eps$; we will call $B$ the dual gauge field. Then the Wilson loop and the wall
 operator combine into a factor
\begin{equation}\label{wallplusregWilson4d}
\exp\left(-\frac{i}{2\pi} \int_{\partial\Z_\eps} A (dB- \Omega_n)\right).
\end{equation}

At this point a remark is in order. The 2-form $\Omega_n$ may not be
unique. This is easy to see from the way it was constructed above:
the relative 2-cycle $D$ is defined modulo an addition of an
absolute 2-cycle. We may fix this ambiguity by requiring the flux of
$\Omega_n$ through an absolute cycle $c_i$ to be $2\pi m_i$ for some
integer $m_i$. What does this freedom correspond to? Examining the
boundary conditions associated with the boundary
action~\eqref{wallplusregWilson4d}, one can easily check that a
saddle point for such a theory -- or, equivalently, a solution to
the classical field equations -- exists only if
$\int_{S^2_\gamma}dB=2\pi n$, i.~e.~if its flux through $S^2_\gamma$
equals that of $\Omega_n$. (One obtains this by integrating the
boundary conditions over $S^2_\gamma$.) This means that topological
sectors not satisfying this flux condition do not contribute to the
path integral. In each topological sector that does contribute, the
curvature of $B$ can be written as
\begin{equation}\label{eq:curvatureSplit4d} F_B=\Omega_n+dB'.
\end{equation}
Here $B'$ is a connection such that $dB'$ has vanishing flux through
$S^2_\gamma$, and $\Omega_n$ may be regarded as the curvature of
some connection on a fixed $U(1)$ bundle. Thus the ambiguity in the
choice of $\Omega_n$ can be regarded as an ambiguity in the
decomposition of $dB$ into a classical background $\Omega_n$ and a
fluctuating part $dB'$. Instead of fixing $\Omega_n$, one might sum
over all possible fluxes of $\Omega_n$ through absolute cycles; then
$B'$ is a connection on a trivial $U(1)$ bundle.

We now arrive at the final stage of the argument. Choosing in each
topological sector an appropriate $\Omega_n$ in the sense just
described, we make use of~\eqref{eq:curvatureSplit4d} and shift the
integration variable in the path-integral:
$$
B=B'+B_n,
$$
where $B_n$ is some connection such that $dB_n=\Omega_n$. In terms
of $B'$ the wall operator \eqref{wallplusregWilson4d} becomes simply
$$
\exp\left(-\frac{i}{2\pi} \int_W A\wedge dB'\right).
$$
The action then splits into parts involving only $B'$ and only $B_n$:
\begin{equation}\label{eq:actionSplits}
S[B;\,\tilde e,\tilde\theta]=S[B';\,\tilde e,\tilde\theta]+
S[B_n;\,\tilde e,\tilde\theta],
\end{equation}
where $\tilde e,\,\tilde\theta$ are the coupling constants of the dual theory.
Indeed, the Maxwell part of the action splits because $dB_n=\Omega_n$ is harmonic,
and the $\theta$-term does not see the topologically trivial field $B'$:
$$
\int d(B_n+B')\wedge d(B_n+B')=\int dB_n\wedge dB_n.
$$

In~\eqref{eq:actionSplits}, the second term does not depend on $B'$ and provides an overall factor
in the path integral. Apart from this factor the
path integral now has exactly the same form as in the absence of a
Wilson loop, i.~e. it has a duality wall inserted at $\partial
\Z_\eps$ and nothing else. Sending $\eps$ to zero we are left with a
path-integral for $B'$ with the standard Maxwell action $S[B';\,\tilde e,\tilde\theta]$. There is also
an overall factor
$$
\exp(-S[B_n;\,\tilde e,\tilde\theta])=\exp\left(-\frac{1}{2\tilde e^2}\int_M dB_n\wedge * dB_n-
\frac{i\tilde\theta}{8\pi^2}\int_M dB_n\wedge dB_n\right)
$$
Recombining it with the action for $B'$ we get the expected
action for a field composed of a topologically trivial part $B'$ and a topological representative
$B_n$ with a singularity on $\gamma$:
$$
S[B_n+B';\,\tilde e,\tilde\theta].
$$
We may therefore say that the dual of $W_n$ is the prescription to
integrate over all fields $B$ that are singular at $\gamma$ so that
the integral of their curvature over the sphere $S^2_\gamma$ equals $2\pi n$. This is the 't~Hooft line operator in the theory of
the field $B$.

\subsubsection{Chern-Simons operator}
Another interesting nonlocal operator is the Chern-Simons operator
$$
\exp\left(-\frac{ik}{4\pi}\int_W A\,dA\right)
$$
It is an operator localized at an oriented connected 3-manifold
$W\subset M$. In the traditional approach, it is not obvious how to
find the $S$-dual of such a nonlocal operator, because it cannot not
be written as a function of $F=dA$ alone.

In the duality wall approach, we consider the tubular neighborhood
of $W$ which is isomorphic to $W\times [-1,1]$ and place the
$S$-wall at $W\times p_1$ and the parity-reversed $S$-wall at
$W\times p_{-1}$, where $p_1$ and $p_{-1}$ are the right and left
boundaries of the interval, respectively. Let us denote $W_-=W\times
p_{-1}$ and $W_+=W\times p_1$. We may regard $W_+$ and $W_-$ as
submanifolds of $M$ (the connected components of the boundary of the
tubular neighborhood of $W$). We rename the gauge field living
outside the tubular neighborhood $A$, and its restrictions to $W_+$
and $W_-$ will be denoted $A_+$ and $A_-$, respectively. In the
limit when the width of the tubular neighborhood goes to zero, the
gauge field $a$ living on $W\times [-1,1]$ becomes
three-dimensional, with an action
\begin{equation}\label{STSwall}
\frac{i}{2\pi} \int_W \left(\frac{k}{2} a\,da +a\,dA_+ -a\,dA_-\right)
\end{equation}
Thus the $S$-dual of the Chern-Simons operator localized on $W$ is a
disorder operator defined by coupling the $U(1)$ Chern-Simons theory
on $W$ to the bulk fields in the way indicated in (\ref{STSwall}).
The traditional approach to $S$-duality runs into trouble in the
case of the Chern-Simons operator because its dual involves a new
field living on $W$.

Since the action for the wall gauge field $a$ is Gaussian, one could
try to integrate it out by solving its equations of motion and
substituting back into the action. The equation of motion reads
$$
k\,da=d(A_--A_+).
$$
Naively, one could therefore replace $a$ with $\frac{1}{k}(A_--A_+)$
and get the following effective action:
$$
-\frac{i}{2\pi}\int_W \frac{1}{2k} (A_+-A_-)d(A_+-A_-).
$$
This is the Chern-Simons action with fractional level $-1/k$.  On
the quantum level this is problematic because the resulting wall
operator is not invariant under large gauge transformation. To avoid
this difficulty, one has to keep the wall field $a$, even though it
only has topological degrees of freedom.

Note that the Chern-Simons operator is nothing but the wall operator
corresponding to the duality transformation
$$
T^k=\begin{pmatrix} 1 & k \\ 0 & 1 \end{pmatrix}.
$$
Its $S$-dual is the wall operator corresponding to the
transformation
$$
S^{-1}T^k S=\begin{pmatrix} 1 & 0 \\ -k & 1\end{pmatrix}
$$

\section{Abelian duality in 2d}

\subsection{Toroidal sigma-model}

The 2d field theory we are going to study is the sigma-model whose
target is a torus $X=T^N\simeq \RR^N/\ZZ^N$ with a flat metric $g$
and a constant B-field $B\in\Omega^2(X)$. The map from the
worldsheet $M$ to $X$ can be thought of as an $N$-component scalar
field $\Phi^I$, $I=1,\ldots,N$ with identifications $\Phi^I\simeq
\Phi^I+2\pi v^I$, where $v^I$ is an arbitrary element of $\ZZ^N$.
The Euclidean action is
$$
S=\frac{1}{4\pi}\int_M \left(g_{IJ} d\Phi^I\wedge * d\Phi^J+ i
B_{IJ}d\Phi^I\wedge d\Phi^J\right).
$$
This class of sigma-models is known to be acted upon by the duality
group $Sp(2n,\ZZ)$. Theories related by this action are equivalent
on the quantum level. The most obvious duality transformation is a
shift of a B-field:
$$
B\mapsto B+8\pi^2 \beta,
$$
where $\beta$ is any 2-form on $X$ with integral periods. Such a
shift modifies the action by a topological term which is an integer
times $2\pi i$, and therefore does not affect the path-integral.
B-field shifts are analogous to shifts of the theta-angle in the 4d
gauge theory.

Another class of duality transformations is T-duality. We will only
consider the simplest T-duality which replaces $X$ with the dual
torus $\hX$. The metric and the B-field on $\hX$ are given by
\begin{equation}\label{TdualmetricBfield}
\hg+\hB=(g+B)^{-1}.
\end{equation}
We will call this the basic T-duality. More general T-dualities
correspond to decomposing $X$ into a product of two tori and
dualizing one of the factors.

First let us exhibit the duality wall corresponding to the B-field
shift. It is a line operator supported on a one-dimensional
submanifold $\Gamma\in M$ and given by
\begin{equation}\label{wallB}
\exp \left(-2\pi i\int_\Gamma \beta_{IJ} \Phi^I d\Phi^J\right).
\end{equation}
Since the fields $\Phi^I$ are defined only up to integer multiples
of $2\pi$, care is required in interpreting this expression. By
analogy with the usual Chern-Simons action, one could try to define
it by picking a Riemann surface $\Sigma$ with
$\partial\Sigma=\Gamma$, extending the map $\Phi:\partial\Sigma\ra
X$ to a map $\tilde\Phi:\Sigma\ra X$, and defining the formal
expression (\ref{wallB}) as
$$
\exp\left(-i\int_\Sigma\tilde\Phi^*\beta \right).
$$
It is easy to see that it does not depend on the choice of $\Sigma$
and $\tilde\Phi$. However, unlike the case of vector bundles, here
an extension $(\Sigma,\tilde\Phi)$ exists only under a very
restrictive condition: the homology class $\Phi_*(\Gamma)\in H_1(X)$
must be trivial. Since $X$ is a torus, if this condition is
satisfied, $\Phi$ can be lifted to a map $\Gamma\ra \RR^N$, and
there is no difficulty in interpreting the expression (\ref{wallB})
anyway.

A better approach is to regard $\beta$ as the curvature 2-form of a
unitary connection $\nabla_\beta$ on a line bundle on $X$. Then one
can define (\ref{wallB}) as the holonomy of the pull-back connection
$\Phi^*\nabla_\beta$. The resulting wall operator is well-defined,
but it depends on the choice of $\nabla_\beta$, not just on $\beta$.
This dependence on $\nabla_\beta$ will show up when we consider the
action of the duality wall on boundary conditions.

Note that if $\Gamma$ separates the worldsheet $M$ into two
disconnected domains, and there are no disorder operator insertions
on one or both of these domains, the map $\Phi$ is necessarily
homologically trivial, and the definition using the holonomy of a
connection reduces to the naive one.

The duality wall corresponding to the basic T-duality is a disorder
line operator separating parts of the worldsheet $M$ where the
sigma-models with targets $X$ and $\hX$ live. Let $\hPhi_I$ be the
fields dual to $\Phi^I$. Then the wall corresponding to the basic
T-duality is
\begin{equation}\label{wallbasicT}
\exp\left(-\frac{i}{2\pi} \int_\Gamma \hPhi_I d\Phi^I \right).
\end{equation}
Again some care is needed in interpreting this expression. Recall
that $\hX$ is, by definition, the moduli space of flat $U(1)$
connections on $X$, and that the connection corresponding to a point
$p=(\hPhi^1,\ldots,\hPhi^N)\in \hX$ is
$$
\frac{1}{2\pi} \hPhi_I d\Phi^I.
$$
One can regard this connection on $X$ as the restriction of the
canonical connection on the Poincar\'{e} line bundle on $X\times
\hX$ to $X\times \{\hat p\}$, where $\hat p$ has coordinates
$\hPhi_J$. The curvature of this connection is
$$
-\frac{1}{2\pi} d\Phi^I\wedge d\hPhi_I .
$$
Then it is natural to define the T-duality wall as the holonomy of
the pull-back of the connection on the Poincar\'{e} line bundle with
respect to the map $(\Phi,\hPhi):\Gamma\ra X\times \hX$.

Varying the action in the presence of the T-duality wall and
requiring the vanishing of the boundary terms we get the following
matching conditions on $\Gamma$:

\begin{equation}\label{eq:matchingCond2d}
\begin{aligned}
\left.\left(g_{IJ}*d\Phi^J+iB_{IJ}d\Phi^J+id\hat\Phi_I\right)\right|_{\Gamma}&=0\\
\left.\left(\hat g^{IJ}*d\hat\Phi_J+i\hat B^{IJ}d\hat\Phi_J+id\Phi^I\right)\right|_{\Gamma}&=0\\
\end{aligned}
\end{equation}
Requiring that the stress-energy tensors match on $\Gamma$ then
gives the relation (\ref{TdualmetricBfield}) between $\hg,\hB$ and
$g,B$. To see this, we introduce the notations $\sigma^{0,1}$ and
$h_{\alpha\beta}$ for the local coordinates and metric on $M$,
denote $F^I_\alpha\equiv \frac{\partial \Phi^I}{\partial
\sigma^\alpha}$ and write the stress-energy tensor as
$$
T^{\alpha\beta}=\frac 1 {4\pi}g_{IJ}F^I_\gamma F^J_\delta
\left(\frac12
h^{\gamma\delta}h^{\alpha\beta}-h^{\alpha\gamma}h^{\beta\delta}
\right)
$$
To simplify the calculations, note that the metric $h^{ij}$ can be
turned into $\delta^{ij}$ by a local diffeomorphism and that without
loss of generality we can assume the ``wall'' $\Gamma$ to be locally
given by the equation $\sigma^1=0$. We can then use the matching
conditions~\eqref{eq:matchingCond2d} to express $\hat F_I$ through
$F^J$ at the wall and verify that of the two independent components
of $T^{\alpha\beta}$, $T^0_1=\hat T^0_1$ identically, while
requiring $T^0_0=\hat T^0_0$ gives the
condition~\eqref{TdualmetricBfield}.

\subsection{Boundary conditions for the sigma-model}

A natural class of boundary conditions for the toroidal sigma-model
is written by analogy with the 4d case. We introduce $n$ boundary
scalars $\phi^i, i=1,\ldots,n,$ with identifications $\phi^i\sim
\phi^i+2\pi v^i$, $v\in\ZZ^n$, and an action
\begin{equation}\label{CSaction0form}
S_b=\frac{i}{2\pi}\int_\gamma \left[\frac12 \kappa_{jk}\phi^j d
\phi^k + \rho_{Jj} \Phi^J d\phi^j+\frac12 \nu_{JK} \Phi^J
d\Phi^K\right],
\end{equation}
where $\gamma$ is a connected component of $\partial M$. Unlike in
4d, here the matrices $\kappa_{ij}$ and $\nu_{JK}$ are
anti-symmetric rather than symmetric. For the boundary theory to be
well-defined, $\kappa$ has to be nodegenerate, and then we may
regard it as a symplectic form on the torus $Y\simeq \RR^n/\ZZ^n$
parameterized by $\phi^i$. Later we will relax the nondegeneracy
constraint.

Since $\phi^i$ are defined up to addition of integer multiples of
$2\pi$, the expression (\ref{CSaction0form}) requires an
interpretation. As above, we introduce a unitary line bundle $\cL$
on $Y\times X$ and equip it with a connection $\nabla$ whose
curvature is
\begin{equation}\label{curv2d}
\frac{1}{2\pi}\left(\frac12\kappa_{jk} d\phi^j d\phi^k+ \rho_{Jj}
d\Phi^J d\phi^j + \frac12 \nu_{JK} d\Phi^J
d\Phi^K\right)=\frac{1}{2\pi} (\kappa+\rho+\nu).
\end{equation}
For this to make sense, the periods of the curvature 2-form must be
$2\pi$ times an integer, and this requires the matrices
$\kappa,\rho,\nu$ to be integral. The choice of $\kappa,\rho,\nu$
determines the topology of the line bundle $\cL$, but leaves some
freedom in the choice of $\nabla$. Any two connections with
curvature (\ref{curv2d}) differ by a flat connection.

In the presence of the boundary-bulk coupling the nondegeneracy
constraint on $\kappa$ can be relaxed. Indeed, if the boundary
fields in the kernel of $\kappa$ all couple nontrivially to the bulk
fields $\Phi^J$, then they can be regarded as Lagrange multiplier
fields whose equations of motion enforce Dirichlet boundary
conditions for some linear combinations of $\Phi^J$. In other words,
the boundary condition is well-defined, if $\rho$, regarded as a map
from $\RR^n$ to $\RR^N$, is injective when restricted to the kernel
of $\kappa$: $(\ker\kappa)\cap(\ker\rho)=0$. A boundary condition satisfying this constraint will be
called nondegenerate.

\subsection{D-brane interpretation}

Let us now explain the interpretation of this class of boundary
conditions in terms of D-branes. Let us first assume that $\kappa$
is nondegenerate. The first step is to quantize the boundary degrees
of freedom regarding the bulk fields $\Phi^J$ as a classical
background. That is, we fix a point $p\in X$ and take $\Phi^J$ to be
the coordinates of this point. Quantization of the boundary fields
$\phi^j$ is a standard exercise in geometric quantization. The torus
$Y$ carries a symplectic form $\kappa$ whose periods are integral
multiples of $(2\pi)^2$. Thus there exists a unitary line bundle
$\cL_p$ on $Y$ with a connection $\nabla_p$ whose curvature is
$\kappa/(2\pi)$. While the topology of the line bundle $\cL_p$ is
determined by $\kappa$, the connection is defined only up to
addition of a closed 1-form. Thus we have to make a choice. This is
the same choice we had to make above when giving the precise meaning
to $\exp(-S_b)$. The bundle $\cL_p$ and the connection $\nabla_p$
are simply the restrictions of $\cL$ and $\nabla$ to the submanifold
$Y\times \{p\}$.

To perform quantization, one also needs to choose a complex
structure on $Y$ so that $\kappa$ is a K\"ahler form, i.~e. of type
$(1,1)$ and positive. This is always possible to do. Then the
quantum-mechanical Hilbert space is the space of holomorphic
sections of $\cL_p$. Its dimension is given by the Riemann-Roch
formula:
$$
h^0(Y,\cL_p)= \int_Y \exp\left(\frac{\kappa}{(2\pi)^2}\right)={\rm
Pf}(\kappa).
$$
{}From the point of view of D-branes, this means that we have a
D-brane on $X$ of rank ${\rm Pf}(\kappa)$.

It is helpful to think about the problem of quantization of the
periodic scalars $\phi^j$ in more physical terms. Consider a
nonrelativistic charged spinless particle on the torus $Y$. We
assume that there is electromagnetic field on $Y$ whose
field-strength is constant and given by $\frac{1}{2\pi}\kappa$. In
this situation the spectrum of the Hamiltonian is discrete; the
energy eigenspaces are called Landau levels. In the limit when the
mass of the particle goes to infinity, the spacing between Landau
levels becomes infinite, while the classical action describing the
particle reduces precisely to (\ref{CSaction0form}). Thus the
quantum theory corresponding to the action (\ref{CSaction0form})
describes the lowest Landau level. Holomorphic sections of the line
bundle $\cL_p$ are the wavefunctions for the lowest Landau level.
Note that in the case when $\dim Y=2$, the formula for
$h^0(Y,\cL_p)$ reduces to
$$
h^0(Y,\cL_p)=\kappa_{12}.
$$
This is the familiar statement that the degeneracy of the Landau
level is proportional to the magnetic field.

As one varies $p\in X$, the connection $\nabla_p$ changes, but the
dimension of $H^0(Y,\cL)$ does not. In this way we obtain a vector
bundle $E$ over $X$. It carries a natural connection: the Berry
connection \cite{Berry,Simon}. In mathematical terms, the Berry
connection on $E$ arises as follows. $E$ is a subbundle of the
infinite-dimensional bundle $\mathcal E$ over $X$ whose fiber over
$p\in X$ is the space of all smooth sections of $\cL_p$. The
connection $\nabla$ on $\cL$ gives rise to a connection on $\mathcal
E$. This in turn induces a connection on the subbundle $E$ via the
orthogonal projection. The evolution operator on the boundary
Hilbert space is simply the holonomy of the Berry connection.

By definition, the charge of a D-brane corresponding to a vector
bundle $E$ on $X$ is the Chern character of $E$. To compute it, note
that the space of holomorphic sections of a positive line bundle on
the complex torus $Y$ can be identified with the kernel of the Dirac
operator on $Y$ twisted by this line bundle. Thus the Chern
character of $E$ can be computed from the family version of the
Atiyah-Singer index theorem:
$$
\ch(E)= \int_Y \exp\left(\frac{F}{2\pi}\right),
$$
where the curvature $F$ is given by (\ref{curv2d}).

The simplest nontrivial example is $X=Y=T^2$. Then the matrices
$\kappa$ and $\nu$ have the form
$$
\kappa_{jk}=\eps_{jk} \kappa_0,\quad \nu_{jk}=\eps_{jk}\nu_0,
$$
where $\eps_{jk}$ is the antisymmetric tensor with $\eps_{12}=1$.
The bundle $E$ has rank $|\kappa_0|$ and its Chern character is
$$
\ch(E)=
\kappa_0+\frac{1}{(2 \pi)^2}(\kappa_0\nu_0-\det\rho)\,d\Phi^1\,d\Phi^2,
$$where $\det\rho\equiv \rho_{11}\rho_{22}-\rho_{12}\rho_{21}$

If $\kappa$ has a nontrivial kernel, the fields $\phi^j$ which
parameterize the kernel are the Lagrange multiplier fields.
Performing the path-integral over these fields puts linear
constraints on the boundary values of the fields $\Phi^J$. These
constraints are
$$
b^j_\alpha\rho_{Jj} \Phi^J=0\ \mod 2\pi .
$$
where the columns of the integral matrix $||b^j_\alpha||$ generate
the kernel of $\kappa$. The constraints define a linear submanifold
$Z\subset X$, and quantization of the remaining fields $\phi^j$ yields
a vector bundle $E$ over $Z$. The pair $(Z,E)$ defines a D-brane on
$X$.

Let us compute the charge of this D-brane. By definition, it is a
cohomology class on $X$ which is the push-forward of the Chern
character of $E$ with respect to the embedding $Z\hookrightarrow X$.
First let us assume that $\kappa=0$, so that all boundary gauge
fields are the Lagrange multiplier fields. In this special case $Z$
is given by the equations
$$
\exp(if_j)=1,\quad j=0,\ldots,n,\quad
$$
where
$$
f_j=\rho_{Jj}\Phi^J.
$$
The bundle $E$ is the trivial bundle of rank one over $Z$. The
charge is simply the Poincar\'{e} dual of the homology class of $Z$.
Up to a sign, it can be represented by a distributional $n$-form
\begin{equation}\label{poincareform}
\prod_{j=1}^n \tilde\delta(f_j) df_j,
\end{equation}
where we regard $f_j$ as a map from $X$ to $\RR/2\pi\ZZ$ and
$\tilde\delta(x)$ is a $\delta$-function on $\RR/2\pi\ZZ$
concentrated at $0$. The precise sign depends on the choice of
orientation for $Z$; we will fix it later.

We can smooth out the form (\ref{poincareform}) without changing its
cohomology class by replacing the distribution $\tilde\delta$ with
any function on $\RR/2\pi\ZZ$ which integrates to $1$, such as the
constant $\frac{1}{2\pi}$. Then the smoothed-out form becomes
$$
(2\pi)^{-n} df_1\ldots df_n .
$$
Up to a sign, we can rewrite it in a more suggestive way as an
integral over $Y$ of an inhomogeneous form on $Y\times X$:
$$
\int_Y \exp\left( \frac{\rho}{(2\pi)^2}\right),
$$
where
$$
\rho=df_j\,d\phi^j=\rho_{Jj} d\Phi^J d\phi^j,
$$
and $\phi^j,j=1,\ldots,n$ are regarded as $2\pi$-periodic
coordinates on $Y$.

If $\kappa\neq 0$, $Y$ can be decomposed into a product of two tori,
on the first of which $\kappa$ is nondegenerate, and the second one
parameterized by Lagrange multiplier fields. Accordingly, the charge
of the D-brane is a product of two contributions which can be
computed as above. They nicely combine into a single integral over
$Y$, so the final result for the charge of the D-brane is simply the
cohomology class of the form
$$
\ch(Z,E)=\int_Y \exp\frac{\kappa+\rho+\nu}{(2\pi)^2}.
$$
We implicitly fixed the orientation convention for $Z$ by requiring
that this formula is valid without any additional sign factors
whether $\kappa$ is nondegenerate or not.

\subsection{Duality action on boundary conditions}

Let us now consider the action of line operators representing
dualities on boundary conditions we have introduced above. To this
end one needs to merge the line operator representing the duality
transformation with the boundary.

For the line operator corresponding to the shift of the B-field this
is very simple: one simply adds to the boundary action a new term
which depends only on the fields $\Phi^J$. As explained above, it is
best to think about this line operator as the holonomy of a certain
connection $\nabla_\beta$ on a line bundle $\cL_{\beta}$ on $X$.
Since the exponential of the boundary action is defined as the
holonomy of a connection $\nabla$ on a line bundle on $Y\times X$,
it is clear that the effect of the B-field shift is simply to add a
new piece to the connection 1-form on $Y\times X$.

Adding a 1-form to a connection gives another connection on the same
line bundle. In our case, we are adding not a globally-defined
1-form, but a pull-back to $Y\times X$ of a connection 1-form on the
line bundle $\cL_\beta$. The result is a connection on a line bundle
on $Y\times X$ which is a product of $\cL$ and the pull-back of
$\cL_\beta$. In other words, the effect of the B-field shift on the
boundary condition is
$$
\cL\mapsto \cL\otimes \pi_X^*\cL_\beta,
$$
where $\pi_X:Y\times X\ra X$ is the projection $(y,x)\mapsto x$.

Now let us consider the basic T-duality. In the limit when the
disorder line operator (\ref{wallbasicT}) merges with the boundary,
the fields $\Phi^J$ living in the gap between the boundary and the
line operator become boundary fields. The new target space for the
boundary degrees of freedom is $Y\times X$. The new boundary action
is
$$
\frac{i}{2\pi}\int_\gamma\left[\frac12 \kappa_{jl} \phi^j d\phi^l +
\rho_{Jj} \Phi^J d\phi^j +\frac12 \nu_{JK} \Phi^J
d\Phi^K + \hPhi_J d\Phi^J\right].
$$
This can be described in words as follows. Given a 2d sigma-model
with target $X$, the boundary condition is specified by a torus $Y$
and a unitary line bundle $\cL$ on $Y\times X$ with a
constant-curvature connection $\nabla$. The basic T-duality has the
following effect:
$$
X\mapsto \hX,\quad Y\mapsto Y\times X,\quad \cL\mapsto \cL\otimes
\pi^* {\mathcal P},
$$
where $\mathcal P$ is the Poincar\'{e} line bundle on $X\times\hX$
and $\pi$ is the projection $Y\times X\times \hX\ra X\times \hX$.

These manipulations are similar to the Fourier-Mukai transform in
algebraic geometry. There, one is given a pair of algebraic
varieties $X$ and $\hX$ as well as an object $B$ of the derived
category of coherent sheaves on $X\times \hX$. For simplicity one
can think about the special case when $B$ is a holomorphic vector
bundle on $X\times\hX$. The object $B$ defines a functor from the
derived category of $X$ denoted $D(X)$ to the derived category of
$\hX$ as follows. Given an object of $D(X)$ one pulls it back to
$X\times \hX$, tensors with $B$ and pushes forward (computes
fiberwise cohomology) to $\hX$. In our case, the role of objects of
the derived category of $X$ is played by trivial torus fibrations
over $X$ with line bundles on them. One can obviously tensor them
over $X$. If the base of the fibration $X$ is decomposed into a
product of two subtori $X_1$ and $X_2$, one can consider the
projection map $\pi_2:X\ra X_2$ and define the push-forward of a
fibration $\mathcal Y$ over $X$ as the same fibration $\mathcal Y$
but regarded as a trivial torus fibration over $X_2$.

If the original boundary condition is nondegenerate, then so should
be the T-dual boundary condition. To see this, we must check that
the condition $(\ker\kappa)\cap(\ker \rho)=0$ is preserved under the
transformation of the triplet $(\kappa,\rho,\nu)$ induced by
$T$-duality (compare with~\eqref{eq:dataTransform4d}):
$$\left\{
\begin{aligned}
\kappa_{n\times n} \quad &\mapsto \quad \tilde \kappa_{(n+N)\times(n+N)}
&&=\begin{pmatrix} \kappa & \rho^t \\ \rho & \nu \end{pmatrix},&&\\
\rho_{N\times n} \quad &\mapsto \quad \tilde \rho_{N\times(n+N)}
&&=(\mathbf 0_{N\times n},\mathbf{1}_{N\times N}),&&\\
\nu_{N\times N} \quad &\mapsto \quad \tilde \nu_{N\times N}
&&= \mathbf 0_{N\times N}.&&
\end{aligned}\right.$$
Here the subscripts indicate the dimensions of the matrices. A vector $\tilde u\in \ker\tilde\rho$
is necessarily of the form
$\left(\begin{smallmatrix}u_n\\\mathbf 0_N\end{smallmatrix}\right)$, and we have
$$
\tilde u\in\ker \tilde \kappa \quad\Leftrightarrow\quad
\begin{pmatrix} \kappa & \rho^t \\ \rho & \nu \end{pmatrix}
\begin{pmatrix}u\\\mathbf 0\end{pmatrix}=
\begin{pmatrix}\kappa u\\
\mathbf \rho u\end{pmatrix}=0\quad\Leftrightarrow\quad u=0,\quad \textsc{Q.E.D.}
$$
Thus the set of boundary conditions we have introduced is closed
both with respect to the basic T-duality and the B-field shifts.

Now it is easy to compute the action of the basic T-duality on
D-brane charges. As explained above, the charge of a D-brane $(Z,E)$
corresponding to the triple $(X,Y,\cL)$ is the integral over $Y$ of
$\ch(\cL)$. The charge of the T-dual D-brane is the integral of
$\ch(\cL\otimes \pi^*\cP)$ over $Y\times X$. Since the Chern
character is multiplicative with respect to the tensor product of
line bundles, this is the same as the integral over $Y\times X$ of
the product of $\ch(\cL)$ and $\pi^*\ch(\cP)$. But this is the same
as the integral over $X$ of the product of $\ch(Z,E)$ and
$\ch(\cP)$. Thus T-duality has the following effect on the D-brane
charge:
$$
\ch(Z,E)\mapsto \ch(\hat Z,\hat E)=\int_X \ch(Z,E) \ch(\cP).
$$
Note that this formula does not require us to know how to realize
any particular D-brane charge on $X$ by a line bundle on $Y\times
X$.

To illustrate this transformation law, let us consider a simple example with $n=2$,
$\kappa_{ij}=r \eps_{ij}$, $\nu_{IJ}=c\eps_{IJ}$ and
$\rho=0$. This corresponds to a D-brane on $T^2$ which has rank $r$ and
first Chern class $cr$, and its charge is
\begin{equation}\label{chargeT2}
r+\frac{cr}{(2\pi)^2} d\Phi^1 d\Phi^2.
\end{equation}
The T-dual boundary condition has $n=4$, ${\hat Y}=Y\times X$ and
the line bundle $\hL$ on $Y\times X\times \hX$ whose first Chern
class is
$$
\frac{1}{(2\pi)^2}\left(r d\phi^1 d\phi^2+c\,d\Phi^1 d\Phi^2 +
d\hPhi_I d\Phi^I\right).
$$
The charge of the corresponding (T-dual) brane on $\hX$ can be computed by
multiplying (\ref{chargeT2}) by the Chern character of the
Poincar\'{e} line bundle and integrating over $X$. The result is
$$
\int_X\left(r+cr\frac{d\Phi^1}{2\pi}\frac{d\Phi^2}{2\pi}\right)
\left(1+\frac{d\hPhi_I}{2\pi}\frac{d\Phi^I}{2\pi}+\frac12
\left[\frac{d\hPhi_I}{2\pi}\frac{d\Phi^I}{2\pi}\right]^2
\right)
=cr-\frac r{(2\pi)^2}d\hPhi_1 d\hPhi_2
$$
Thus T-duality essentially exchanges the rank and the first Chern
class of the D-brane.

\section{Abelian duality in 3d}

\subsection{The duality wall in 3d}

Consider a $U(1)$ gauge theory on a Riemannian 3-manifold $M$ with a
Euclidean action
$$
S=\frac{1}{2e^2}\int_M F\wedge *F.
$$
This theory is equivalent to a bosonic sigma-model on $M$ with
target $S^1$. The basic field of this sigma-model is a scalar
$\sigma$ with the identification $\sigma\sim \sigma+2\pi$ and an
action
$$
\frac{e^2}{8\pi^2} \int_M  d\sigma\wedge * d\sigma.
$$
The corresponding duality wall is given by an insertion of an
operator
\begin{equation}\label{wall3d}
\exp\left(\frac{-i}{2\pi}\int_W A\,d\sigma\right),
\end{equation}
where $W$ is an oriented surface in $M$ splitting it into $M_+$ and
$M_-$. By convention, the orientation of $W$ agrees with that of
$M_-$ and disagrees with that of $M_+$. The fields $A$ and $\sigma$
are defined on $M_-$ and $M_+$, respectively. Their boundary values
on $W$ are unconstrained. Varying the action and requiring the
vanishing of boundary terms in the variation gives the following
matching conditions on $W$:
\begin{align}\label{match3d1}
*F\vert_W &=\frac{ie^2}{2\pi} d\sigma\vert_W, \\ \label{match3d2}
F\vert_W &= \frac{ie^2}{2\pi} *d\sigma\vert_W .
\end{align}
One can check that these conditions ensure the continuity of the
stress-energy tensor across $W$. This shows that the wall operator is
indeed topological.

The definition of the duality wall given above is somewhat
imprecise, because the connection 1-form $A$ is defined only up to a
gauge transformation. Formal integration by parts does not
ameliorate the situation, because the expression
$$
\int_W F \sigma,
$$
where $\sigma$ is defined up to a multiple of $2\pi$, is also
ill-defined. In fact, integration by parts makes the matters worse,
because it creates an illusion that the wall operator depends only
on $F$, not $A$, which is incorrect.

The most natural interpretation of the expression (\ref{wall3d}) is
in terms of a connection on a $U(1)$ gerbe over $W$. In the string
theory context, a gerbe connection is also known as a B-field. For a
brief review of the topology of the B-field see e.~g.
\cite{gerbehol2}. For our purposes, a gerbe connection is given by a
collection of 2-forms $B_i$ defined on charts $U_i$ of an open cover
of $W$, where $i$ runs over the index set of the cover. The 2-forms
satisfy gluing conditions on the overlaps $U_{ij}=U_i\bigcap U_j$:
$$
B_i-B_j=dA_{ij}.
$$
Here $A_{ij}$ are $U(1)$ connection 1-forms on open sets $U_{ij}$.
They obviously satisfy $A_{ij}=-A_{ji}$. As part of the definition
of the gerbe connection, $A_{ij}$ are required to satisfy a gluing
condition of their own on triple overlaps $U_{ijk}=U_i\bigcap
U_j\bigcap U_k$:
$$
A_{ij}+A_{jk}+A_{ki}=d f_{ijk},
$$
where $f_{ijk}$ are $U(1)$-valued functions on triple overlaps
completely antisymmetric with respect to indices $ijk$. Here we
think about $U(1)$ as $\RR/2\pi\ZZ$ rather than as the unit circle
in the complex plane, to avoid the annoying factors $\sqrt {-1}$ in
the formulas. The functions $f_{ijk}$ must satisfy a cocycle
condition on quadruple overlaps $U_{ijkl}$:
\begin{equation}\label{4cocycle}
f_{ijk}+f_{jkl}+f_{kli}+f_{lij}=0 \ {\rm mod}\, 2\pi\ZZ.
\end{equation}
The gerbe itself is defined by functions $f_{ijk}$; the rest of the
data defines a connection on this gerbe. It should be clear from the
above description that a gerbe is a ``higher analogue'' of a line
bundle.

Just like an ordinary connection on a line bundle assigns an element
of $U(1)$ (the holonomy) to any loop in the manifold, a gerbe
connection assigns an element of $U(1)$ to any closed oriented
two-dimensional submanifold. This number is usually called gerbe
holonomy.\footnote{In string theory context, gerbe holonomy is the
factor in the path-integral arising from the B-field.} In our case,
since $W$ is two-dimensional, the only submanifold of interest is
$W$ itself.

The construction of gerbe holonomy is described in detail in
\cite{gerbehol1}. In brief, one triangulates $W$ and considers an
open cover whose charts are so-called open stars of the
triangulation labeled by vertices of the triangulation. (An open
star associated to a vertex is the interior of the union of all
simplices containing this vertex.) Then one considers a cell
decomposition dual to the triangulation. The 2-cells are labeled by
vertices of the triangulation, 1-cells are labeled by edges of the
triangulation, and 0-cells are labeled by simplices of the
triangulation (they are baricenters of the simplices). Note that
each 2-cell lies entirely within a single open chart and therefore
can be naturally associated with this open chart. Similarly, each
1-cell belongs to the closure of precisely two 2-cells and belongs
to the overlap of the corresponding open charts; therefore it can be
naturally associated to an overlap of two charts. Finally, each
0-cell lies in a unique triple overlap. To put it differently, each
2-form $B_i$ naturally lives on a particular 2-cell, each 1-form
$A_{ij}$ naturally lives on a particular 1-cell, and each $f_{ijk}$
naturally lives on a particular 0-cell.

Now the gerbe holonomy is defined as follows. One integrates all
$B_i$ over the corresponding 2-cells, all $A_{ij}$ over the
corresponding 1-cells, evaluates $f_{ijk}$ on the corresponding
0-cells and adds up the results. This gives an element of
$\RR/2\pi\ZZ$ which one can then multiply by $\sqrt {-1}$ and
exponentiate. The resulting phase is the gerbe holonomy. One can
show that the gerbe holonomy is independent of the choice of the
triangulation.

Now we show that the expression $\frac{1}{2\pi} A d\sigma$ can be
naturally interpreted as a connection on a gerbe. We pick an open
cover of $W$
$$
\left\{U_i,i\in {\mathsf S}\right\},
$$
such that all $U_i$ and all multiple overlaps are contractible.
Recall that the gauge field $A$ is defined by a collection of
1-forms $A_i$ on $U_i$ such that on double overlaps we have
$$
A_i-A_j=d g_{ij},
$$
where $g_{ij}$ are $\RR$-valued transition functions. (The usual
complex-valued transition functions are obtained by multiplying
$g_{ij}$ by $\sqrt {-1}$ and exponentiating.) On triple overlaps
they must satisfy
$$
g_{ij}+g_{jk}+g_{ki}=2\pi m_{ijk},\quad m_{ijk}\in\ZZ.
$$
The integers $m_{ijk}$ form a Cech 2-cocycle representing the first
Chern class of the $U(1)$ gauge bundle. Similarly, one can specify
$\sigma$ by $\RR$-valued functions $\sigma_i$ on $U_i$ which satisfy
on double overlaps
$$
\sigma_i-\sigma_j=2\pi r_{ij},\quad r_{ij}\in \ZZ.
$$
The integers $r_{ij}$ form a Cech 1-cocycle representing the
cohomology class of the 1-form $d\sigma$.

To define a gerbe connection on $W$ we let
$$
B_i=\frac12 \left(A_i \frac{d\sigma_i}{2\pi}+dA_i
\frac{\sigma_i}{2\pi}\right).
$$
Then on double overlaps $U_{ij}$ we have
$$
B_i-B_j=\frac12\left(dg_{ij} \frac{d\sigma_j}{2\pi}+dA_i
r_{ij}\right).
$$
The right-hand side can be written as a total derivative
$$
\frac12 d\left(A_i r_{ij}+\frac{\sigma_j}{2\pi} dg_{ij}+2 g_{ij}
\frac{d\sigma_j}{2\pi}\right).
$$
We therefore let
$$
A_{ij}=\frac12 \left(A_i r_{ij}+\frac{\sigma_j}{2\pi}
dg_{ij}\right)+g_{ij} \frac{d\sigma_j}{2\pi}.
$$
One can check that the 1-forms $A_{ij}$ satisfy $A_{ij}=-A_{ji}$.
One can also check that on triple overlaps we have
$$
A_{ij}+A_{jk}+A_{ki}=m_{ijk} d\sigma_j .
$$
Hence we let
$$
f_{ijk}=m_{ijk} \sigma_j.
$$
This function is well-defined as a map from $U_{ijk}$ to
$\RR/2\pi\ZZ$, is completely antisymmetric with respect to all
indices, and obviously satisfies the cocycle condition
(\ref{4cocycle}) on quadruple overlaps. Hence we have successfully
defined a gerbe on $W$ and a connection on it. The duality wall will
be defined as the holonomy of this gerbe connection.

One can modify the $U(1)$ gauge theory by adding a Chern-Simons term
to the action. This modification destroys the duality between the
gauge theory and the sigma-model. In the duality wall formalism this
comes about from the fact that the Chern-Simons term is not gauge
invariant in the presence of a free boundary on which $F=dA$ is
allowed to be nonzero. Thus the duality wall operator as defined
above is not even gauge-invariant. One could try rectify this by
placing an extra degree of freedom on the wall which couples to the
gauge field in an anomalous manner. For example, one could put a 2d
chiral gauge boson on $W$ which couples to $A$. But the resulting
wall operator is not topological and does not define a duality
transformation.

\subsection{Duality action on operators}

Let us give some examples demonstrating how the duality wall acts on
operators in the dual theories. First of all, the matching
conditions (\ref{match3d1},\ref{match3d2}) imply that the local
operator $F$ is mapped to
$$
\frac{ie^2}{2\pi} * d\sigma.
$$
This is usually taken as the definition of the duality
transformation.

Another local operator is the 't~Hooft operator. In 3d it is a local
operator (i.~e. it is localized at a point $p\in M$). It is a
disorder operator defined by the condition that at the insertion
point $p$ $F$ is singular and satisfies
\begin{equation}\label{flux3d}
\int_W \frac{F}{2\pi}=m.
\end{equation}
Here $W$ is a small 2-sphere centered at $p$. The integer $m$ is
called the magnetic charge. The dual of the 't~Hooft operator is the
operator $\exp(-im\sigma(p))$. To see this, we place the duality
wall at $W$. Since $W$ is simply-connected, the map $\sigma$ can be
thought of as an ordinary $\RR$-valued function and the gerbe
holonomy simplifies to
$$
\exp\left(\frac{-i}{2\pi}\int_W F
\sigma\right)=\exp\left(-im\sigma\right).
$$
Now the integral over $A$ is decoupled from the integral over
$\sigma$, and therefore integration over $A$ produces an irrelevant
constant.

Next we consider the action of duality on line operators in the
gauge theory. The simplest such operator is the Wilson line operator
$$
W_n(\gamma)=\exp\left(in\int_\gamma A\right)
$$
where $\gamma$ is a circle embedded into $M$, and $n\in\ZZ$ is the
electric charge. The argument allowing to calculate its dual is
essentially identical to the four-dimensional case discussed in
section~\ref{sec:dualizingWilson4d}, only simplified because of the
absence of a $\theta$-term in the action. First we are going to
regularize the Wilson operator. Let $\Z_\eps$ be a tubular
neighborhood of $\gamma$ of  ``thickness'' $\eps$. Let $\psi_n$ be a
harmonic map $M\bs \Z_\eps\ra S^1$ satisfying
$$
* d\psi_n\vert_{\partial \Z_\eps}=0\quad \text{ and }\quad
\int_L d\psi_n=2\pi n,
$$
where $L$ is a small circle which has linking number $1$ with
$\gamma$. At this stage we assume that $\gamma$ is homologically
trivial, so that the linking number is well-defined. (As in the 4d
case, if $\gamma$ is homologically nontrivial and there are no other
Wilson loops inserted, the path-integral vanishes identically
because the integrand is not invariant with respect to adding to the
connection a flat connection with a nontrivial holonomy along
$\gamma$). The regularized Wilson loop is defined to be
$$
W^\eps_n(\gamma)=\exp\left(\frac{i}{2\pi}\int_{\partial\Z_\eps} A\wedge
d\psi_n\right).
$$
More generally, if several Wilson loops $\gamma_1,\ldots,\gamma_K$
with charges $n_1,\ldots,n_K$ are present simultaneously, the
path-integral can be nonvanishing only if the following condition is
satisfied:
\begin{equation}\label{eq:homologicalCond}
\sum_{i=1}^K n_i [\gamma_i]=0.
\end{equation}
Here $[\gamma_i]$ denotes the homology class of $\gamma_i$. One can
show then that on the complement of the tubular neighborhood
${\mathcal Z}_\eps$ of all the loops $\gamma_i$ there exists a
harmonic map $\psi_n$ which satisfies
$$*d\psi_n\vert_{\partial{\mathcal Z}_\eps}=0,\quad \int_{L_i} d\psi_n=2\pi n_i,$$
where $L_i$ is a small circle which winds once around $\gamma_i$.

The 1-form $d\psi_n$ can be seen as a magnetic field created by $K$
superconducting wires filling the tubular neighborhoods of
$\gamma_i, i=1,\ldots,K$, such that the $i^{\rm th}$ wire carries
electric current $2\pi n_i$. The multivalued function $\psi_n$ is the
``magnetic potential.''

Having regularized the Wilson operator by transforming it into an integral over $\partial \Z_\eps$,
we place the duality wall on this surface and notice that the Wilson loop and the wall operator combine
into a factor
\begin{equation}\label{wallplusregWilson}
\exp\left(-\frac{i}{2\pi} \int_W A (d\sigma- d\psi_n)\right),
\end{equation}
where $W=\partial\Z_\eps$ is the $S$-duality wall.\footnote{The
gerbe holonomy again simplifies in this case because the $U(1)$
bundle is trivial when restricted to $W$ and the 1-form $A$ can be
thought of as an ordinary 1-form on $W$.} Then, as in
four-dimensional case, we shift the integration variables:
$$
\sigma=\sigma'+\psi_n.
$$
and note that in terms of $\sigma'$ the wall operator
(\ref{wallplusregWilson}) becomes
$$
\exp\left(-\frac{i}{2\pi} \int_W A \,d\sigma'\right),
$$
while the bulk action splits in two parts
$$
\frac{1}{2e^2}\int_M d\sigma'\wedge *d\sigma'+\frac{1}{2e^2} \int_M
d\psi_n\wedge *d\psi_n.
$$
As in the four-dimensional case, we recognize the situation as a
$S$-duality wall separating two theories with no operator
insertions, with weights of different topological sectors modified
by an overall multiplicative factor. We eliminate the field $A$ by
sending $\epsilon$ to zero; then we are left with a path-integral
for $\sigma'$ with the standard action and the multiplicative factor
which we can reabsorb into the action. The new action is
$$
\frac{1}{2e^2}\int_M d(\sigma'+\psi_n)\wedge * d(\sigma'+\psi_n).
$$
Here $\sigma'$ is nonsingular at $\gamma$, while $\psi_n$ has a
singularity of the form
$$
n \phi +\text{{regular terms}}.
$$
Here we identified the neighborhood of $\gamma$ with $S^1\times
D^2$, where $D^2$ is a small disc in $\RR^2$ with a radial
coordinate $r$ and angular coordinate $\phi$. Since the action
depends only on the sum $\sigma=\sigma'+\psi_n$, we may say that the
dual of $W_n$ is the prescription to integrate over all fields
$\sigma$ such that $\sigma$ has the same singularity at $\gamma$ as
$\psi_n$. This defines a disorder operator in the theory of the
field $\sigma$.

There is also a natural disorder line operator in the 3d gauge
theory. It is the reduction of the Gukov-Witten-type surface
operator \cite{GuWi} from 4d to 3d (with the reduced dimension taken
to be along the surface). It is defined by the condition
$$
F=\alpha\delta_\gamma,
$$
where $\delta_\gamma$ is a delta-current supported on $\gamma$.
Another way to define is to say that the holonomy of $A$ along a
small circle linking $\gamma$ approaches $\exp(2\pi i\alpha)$ as the
size of the circle shrinks to zero. The second definition makes it
clear that $\alpha\in \RR$ is defined modulo $\ZZ$.

To find its dual, we again place the duality wall at the boundary
$W$ of the tubular neighborhood $\Z_\eps$ of $\gamma$. Up to regular
terms, the restriction of $A$ to $W$ is given by $\alpha d\psi_1$,
where $\psi_1:M\bs\Z_\eps\ra S^1$ has been defined above. Hence up
to terms which vanish in the limit $\eps\ra 0$ the duality wall
operator becomes
$$
\exp\left(\frac{-i\alpha}{2\pi} \int_W d\psi_1 d\sigma\right).
$$
This is a regularization of
\begin{equation}\label{dualGuWi}
\exp\left(i\alpha\int_\gamma d\sigma\right).
\end{equation}
One can describe this operator in words as follows. Consider a
$U(1)$ gauge field on $S^1$ whose holonomy is $\exp(i\alpha)$. Given
a map $\sigma:M\ra S^1$, we may pull back this $U(1)$ gauge field to
$M$ and evaluate its holonomy along $\gamma\subset M$. The resulting
observable is precisely (\ref{dualGuWi}). One may call this
observable a Wilson line operator; they have been introduced by
Rozansky and Witten in the context of topological sigma-models in 3d
\cite{RW}.

\subsection{Duality action on boundary conditions}

The two most natural boundary conditions in the 3d gauge theory are
the Dirichlet and Neumann boundary conditions. The Dirichlet
condition is analogous to the PEC condition in 4d: we require
$A\vert_{\partial M}$ to be equal to a fixed flat connection on
$\partial M$. Thus $F$ vanishes when restricted to the boundary. The
Neumann boundary condition corresponds to keeping $A\vert_{\partial
M}$ unconstrained and imposing
$$
*F\vert_{\partial M}=0.
$$
On the quantum level the Neumann condition has a modulus: we can add
to the action a boundary term
$$
\frac{i\theta}{2\pi}\int_{\partial M} F.
$$
Clearly, the parameter $\theta$ is defined modulo $2\pi$.

The duality maps the Dirichlet condition in the gauge theory to the
Neumann condition in the sigma-model:
$$
*d\sigma\vert_{\partial M}=0.
$$
This boundary condition has moduli: one can pick an arbitrary flat
connection $a$ on $\partial M$ and add to the action a boundary term
$$
\frac{i}{2\pi}\int_\pM a\,d\sigma
$$
(More precisely, we should understand the corresponding phase factor
in the path-integral as the holonomy of a gerbe.) Using the duality
wall argument one can check that duality maps the flat connection
$A\vert_{\partial M}$ to $-a$. Indeed, as the duality wall merges
with the Neumann boundary in the scalar field theory, the boundary
action becomes
$$
\frac{i}{2\pi}\int_W (A+a) d\sigma .
$$
If we separate $\sigma$ into a topologically-trivial and
topologically-nontrivial parts, integration over the
topologically-trivial part gives a functional delta-function which
forces $F=dA$ to vanish on the boundary, while integration over the
topologically-nontrivial part forces the holonomy of the flat
connection $A+a$ to vanish.

The duality also maps the Neumann condition in the gauge theory to
the Dirichlet condition in the sigma-model:
$$
d\sigma\vert_{\partial M}=0.
$$
This means that $\sigma$ is a constant $\sigma_0$ on any connected
component of $\partial M$. The modulus $\sigma_0$ is dual to the
modulus $\theta$ in the gauge theory. Indeed, as the duality wall
merges with the Dirichlet boundary in the scalar field theory, it
becomes a boundary term in the action
$$
\frac{i}{2\pi}\int_\pM F \sigma_0.
$$

We may also consider more general boundary conditions in the gauge
theory involving boundary degrees of freedom coupled to the gauge
field $A$ in a gauge-invariant manner. For example, we may consider
a sigma-model with target $X$ which admits a $U(1)$ isometry and
promote this symmetry to a gauge symmetry by replacing ordinary
derivatives with covariant derivatives. To determine the dual of
this boundary condition, we place the duality wall infinitesimally
close to the boundary. The gauge field $A$ in the gap between the
wall and the boundary becomes effectively two-dimensional, and the
boundary action becomes
$$
S_{bdry}(\phi,A)+\frac{i}{2\pi}\int_{\partial M} F\sigma
$$
Since the 2d gauge field has no kinetic term, one can regard the
boundary theory as the strong-coupling limit of the ordinary gauged
sigma-model with target $X$, with the 2d theta-angle promoted to a
field $\sigma\vert_{\partial M}$.

\end{document}